\documentclass[sigconf]{acmart}

\AtBeginDocument{%
  \providecommand\BibTeX{{%
    \normalfont B\kern-0.5em{\scshape i\kern-0.25em b}\kern-0.8em\TeX}}}

\setcopyright{acmcopyright}
\copyrightyear{2018}
\acmYear{2018}
\acmDOI{10.1145/1122445.1122456}

\acmConference[Woodstock '18]{Woodstock '18: ACM Symposium on Neural
  Gaze Detection}{June 03--05, 2018}{Woodstock, NY}
\acmBooktitle{Woodstock '18: ACM Symposium on Neural Gaze Detection,
  June 03--05, 2018, Woodstock, NY}
\acmPrice{15.00}
\acmISBN{978-1-4503-XXXX-X/18/06}

\usepackage{float}
\usepackage{amsmath}
\usepackage{bm}
\usepackage{multirow}
\usepackage{graphics}
\usepackage[table,xcdraw]{}

\begin{document}

\setlength{\abovedisplayskip}{3pt}
\setlength{\belowdisplayskip}{3pt}

\newcommand{\EG}{{e.g.}}
\newcommand{\IE}{{i.e.}}
\newcommand{\ETAL}{et al.\ }
\newcommand{\etc}{etc}
\newcommand{\R}{\mathbb{R}}

\title{Individual-level Anxiety Detection and Prediction from Longitudinal YouTube and Google Search Engagement Logs}

\author{Anis Zaman}
\affiliation{\institution{University of Rochester}}
\email{azaman2@cs.rochester.edu}

\author{Boyu Zhang}
\affiliation{\institution{University of Rochester}}
\email{bzhang25@u.rochester.edu}

\author{Henry Kautz}
\affiliation{\institution{University of Rochester}}
\email{henry.kautz@gmail.com}

\author{Vincent Silenzio}
\affiliation{\institution{Rutgers University}}
\email{vincent.silenzio@rutgers.edu}

\author{Ehsan Hoque}
\affiliation{\institution{University of Rochester}}
\email{mehoque@gmail.com}



\begin{abstract}
Anxiety disorder is one of the world's most prevalent mental health conditions, arising from complex interactions of biological and environmental factors and severely interfering one’s ability to lead normal life activities. Current methods for detecting anxiety heavily rely on in-person interviews, which can be expensive, time-consuming, and blocked by social stigmas. 
In this work, we propose an alternative method to identify individuals with anxiety and further estimate their levels of anxiety using personal online activity histories from YouTube and the Google Search engine, platforms that are used by millions of people daily. We ran a longitudinal study and collected multiple rounds of anonymized YouTube and Google Search logs from volunteering participants, along with their clinically validated ground-truth anxiety assessment scores. We then developed explainable features that capture both the temporal and contextual aspects of online behaviors. Using those, we were able to train models that (i) identify individuals having anxiety disorder with an average F1 score of $0.83 \pm 0.09$ and (ii) assess the level of anxiety by predicting the gold standard Generalized Anxiety Disorder 7-item scores (ranges from 0 to 21) with a mean square error of $1.87 \pm 0.15$ based on the ubiquitous individual-level online engagement data. Our proposed anxiety assessment framework is cost-effective, time-saving, scalable, and opens the door for it to be deployed in real-world clinical settings, empowering care providers and therapists to learn about anxiety disorders of patients non-invasively at any moment in time.
\end{abstract}




\maketitle

\section{Introduction}
According to the World Health Organization (WHO), 1 in 13 people suffers from anxiety globally, making it one of the most prevalent mental health concerns. In the United States, it is the second leading cause of disability among all psychiatric disorders~\cite{whiteford2013global}. Nearly 40 million people (age 18 and older) experienced anxiety disorder in any given year, yet only $35.9\%$ of those suffered received treatments\footnote{\url{https://adaa.org/understanding-anxiety}}. A study in 2017 reported that the level of anxiety among young adolescents has been gradually increasing in recent years~\cite{calling2017longitudinal}.

The population most vulnerable to anxiety disorder is the students in high school and early college years. A report by the American College Health Association in 2018 stated that $63\%$ of college students in the U.S. felt overwhelming anxiety during the last 12 months, and only $23\%$ of these students were either diagnosed or treated for an anxiety disorder by a professional mental healthcare provider~\cite{americancollegehealthassociation_2018}. During the early days of college, students are separated from their traditional support system and find themselves in challenging social and academic settings such as living with roommates, developing independent identities, making new friends, managing heavy workloads, \etc. All these experiences induce spikes in anxiety from time to time ~\cite{purdon2001social}, and this psychological distress increases during the first few semesters of college~\cite{conley2020navigating}. Furthermore, it has been reported that anxiety disorders are significantly associated with other medical and psychiatric comorbidities~\cite{costello2005developmental}. Despite such a high prevalence of anxiety among young adolescents, current methods for detecting anxiety disorders consist of self-assessment surveys and in-person interviews, which can be time-consuming, expensive, lack precision, and hampered by factors such as fear, concealing information, and social stigma related to the mental health issue. 

Engagements in online platforms are major components in the lives of young adults~\cite{kaplan2010users}. On average, an internet user spent the equivalent of more than 100 days online during the last 12 months~\cite{TNR2020}. It has been reported that $81\%$ of U.S internet users aging between 15 to 25 use YouTube\footnote{\url{https://www.statista.com/statistics/296227/us-youtube-reach-age-gender/}} regularly. Besides, an average internet user uses Google Search at least once a day, and many search dozens of times a day\footnote{\url{https://bit.ly/382vgWD}}. Extensive studies have been done trying to correlate mental health issues with popular public social media data such as Facebook~\cite{ophir2020if,brailovskaia2019relationships,nisar2019facebook} and Twitter~\cite{coppersmith2014quantifying,chancellor2020methods,coppersmith2014measuring,gopalakrishna2018detection}, yet they may fail to cover people who interact infrequently with social media or post false positive impressions publicly~\cite{FacebookFalse}. In contrast, individual-level search and YouTube logs are ubiquitous and private for each user and are less likely to be subject to self-censorship. A group of researchers has shown that search logs can be used as a proxy for detecting mental health issues~\cite{adler2019search,jimenez2020google,zaman2019detecting}. We draw inspirations from these prior works and hypothesize that private Google Search engine logs and YouTube histories can leave a detailed digital trace of the mental health states of users and be used as a proxy to assess the level of anxiety for individuals.

In this work, we propose a framework that leverages individual-level online activities logs, in particular, Google Search and YouTube activity histories, to \textit{identify} individuals with anxiety disorder and further \textit{predict} their level of anxiety. We ran a longitudinal study to gather two rounds of data, with 5 months in-between, from a college population. During each round, participants shared their anonymized online activity histories along with their answers to a clinically validated questionnaire for measuring Generalized Anxiety Disorder (GAD-7)~\cite{spitzer2006brief}. We then developed an explainable low-dimensional vector representation that captures different aspects of one’s online behaviors, including temporal activity patterns, time and semantic diversities, and periods of inactivity. Using these feature representations, we trained models that can accurately detect and predict one's level of anxiety from online activities. Unlike~\cite{zaman2019detecting} who merely focused on mental health issue \textit{detection} such as self-esteem from Google Search histories, our data incorporates both Google Search as well as YouTube activities history, and our two rounds of data facilitate both the \textit{detection} and \textit{prediction} tasks. Furthermore, we conduct our experiment with a framework that fits possible real-world applications. We envision our work as an important step towards helping caregivers better understand and engage with their patients without additional burden through passive data and ubiquitous computing.

In summary, this work is unique in that (i) we are the first to run a longitudinal study where individual-level Google Search and YouTube histories along with gold-standard clinically validated anxiety assessment are gathered; (ii) we define explainable features that capture both the semantic and temporal aspects of online activities, including a novel representation of periods of activity and inactivity based on temporal point processes; (iii) using these features, we managed to both detect and predict the anxiety disorder of an individual with high performances, showing that ubiquitous private online logs contain strong signals that can potentially be a proxy to assess mental health issues; and (iv) our pioneered two rounds of data and light-weight experiment setup has strong societal implications and can empower care providers to estimate the anxiety levels of patients remotely through a non-invasive manner.

\section{Related Work}
Public social media, blogs, and forums have become popular data sources for researchers to study the prevalence of mental health conditions. ~\cite{seabrook2016social} showed that the usage of social media sites correlates with user depression and anxiety. Tweets, one of the most explored social media platform, has been used to detect insomnia~\cite{jamison2012can}, suicidal ideations ~\cite{de2016discovering}, depressed individuals~\cite{de2013predicting}, the extent of depression ~\cite{tsugawa2015recognizing}, and languages related to depression and PTSD ~\cite{reece2017forecasting,coppersmith2015clpsych,resnik2015beyond,pedersen2015screening}. Besides, ~\cite{de2014characterizing} have shown that Facebook status can be used to predict postpartum depression and monitor depression~\cite{schwartz2014towards}. Other researchers have leveraged data from Reddit to study mental distress among adolescents ~\cite{bagroy2017social}. ~\cite{de2016discovering} identified shifts in language may indicate future suicidal ideations. De Choudhury \ETAL provides a comprehensive overview of the role of social media in mental health researches ~\cite{de2013social} and evaluation methodologies ~\cite{chancellor2020methods}. Social media users constitutes only a fraction of the general population, and a small number of them, with particular personalities or demographics, typically acts out in social media that may reveal signs of mental health struggles. Hence, findings based on social media platforms may not generalize to the majority of the population.

A large number of researchers have leveraged sensors, such as smartphone and mobile apps, that are embedded in our daily life experiences to capture various aspects of mental health~\cite{mohr2017personal,wang2014studentlife}. For example, sensor data has been used for studies of anxiety ~\cite{elhai2019relationship,place2017behavioral,richardson2018problematic}, stress ~\cite{sano2013stress,muaremi2013towards}, moods~\cite{likamwa2013moodscope,ma2012daily}, and
depression ~\cite{wang2018tracking,saeb2015mobile,saeb2016relationship}. Several research groups have developed applications to help users manage stress and anxiety ~\cite{maclean2013moodwings,paredes2011calmmenow} and evoke positive emotions ~\cite{amores2016psychicvr}. However, smartphone applications for tackling mental health issues have several limitations: (i) not every mental health patient have access to smartphones; (ii) any interventions delivered via apps is less likely to be as effective as face-to-face sessions with a therapists ~\cite{zhang2015smartphone}; (iii) the app may fail and require developers to constantly keep it updated, which is costly and not sustainable. 


One data source that can capture in-the-moment thoughts and feelings of a broad range of people are search engine logs, which may fill in the gap for continuous monitoring applications ~\cite{mohr2017personal}. Researchers have used population-level search engine logs from Google Trends to monitor depression and suicide-related behaviors ~\cite{mccarthy2010internet,sueki2011does,yang2011association,gunn2013using}, identify seasonality in seeking mental health information~\cite{ayers2013seasonality}, and show heavy usages for screening diseases ~\cite{paparrizos2016detecting} such as pancreatic cancer~\cite{paparrizos2016screening}. A comprehensive review of the usage of Google Trends in the healthcare domain has been provided by ~\cite{nuti2014use}. A crucial difference between these previous works and ours is that we aim to accurately predict the mental health of particular individuals, not general populations. Unlike population-level online engagement logs in Google Trends, our individual-level activity logs are more likely to fit the fabric of one's daily life experience.

\section{Data}
\begin{figure}
\includegraphics[scale=0.4]{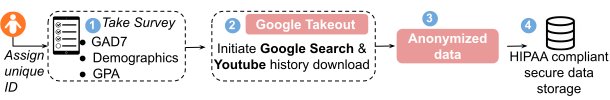}
\centering
\caption{Obtaining data from an individual}
\label{data_download}
\vspace{-4mm}
\end{figure}

The longitudinal data collected for this work consisted of individual-level Google Search logs, YouTube history, and clinical survey responses that are very personal and sensitive in nature. Similar to~\cite{zaman2019detecting}, we leveraged a cloud-based data collection process using Google Takeout\footnote{\url{http://takeout.google.com/}}, a web interface that enables Google product users to export their Google Search and YouTube activity histories. Our cloud-based data collection pipeline (see Figure~\ref{data_download}) has been thoroughly vetted by the Institutional Review Board (IRB) of our institution in order to ensure the privacy and safety of subjects.

\subsection{Study Recruitment Procedure}
\label{recruitment_procedure}
The study ran for 5 months starting in August, 2019. Participation was voluntary, and one needed to be at least 18-year-old and have a Google account to qualify for the study. The recruitment procedure was designed as an one-on-one interview. During the recruitment, participants answered the 7-item Generalized Anxiety Disorder questionnaire, a clinically validated tool for assessing anxiety disorder, in addition to their GPA, gender, and demographics. Following that, participants signed in to Google Takeout with their Google accounts and initiated the Google Search and YouTube activity history data download process. Before the data was shared with the research team, all sensitive information such as name, email, phone number, social security, and financial information (banking and credit card) was redacted and anonymized using Google's Data Loss Prevention (DLP) API~\cite{kiang2016data,kim2016cloud}. 

In total, we collected two rounds of data. The recruitment procedure above was performed during each round. In August 2019, $104$ qualified college college students participated in the first round. For the rest of the paper, we will refer this round of data as the \textit{first-round} data. 

\begin{figure}
\includegraphics[scale=0.32]{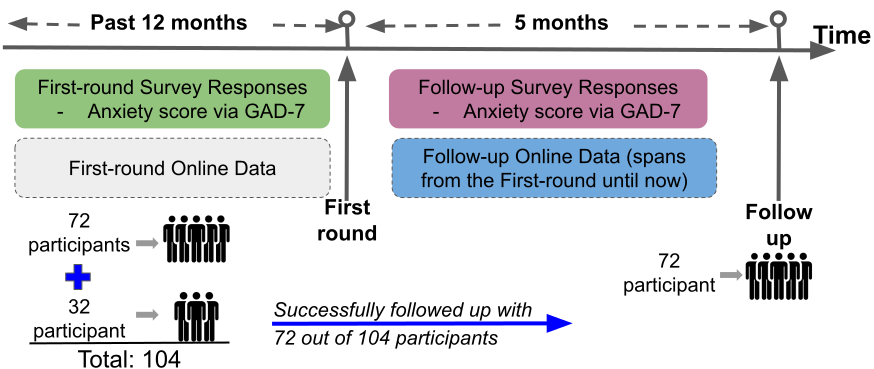}
\centering
\caption{Timeline and participants for two rounds of data collection. There are in total $104$ unique individuals participated in the study, and $72$ of them participated in both the ~\textit{first-round} and the ~\textit{follow-up}.}
\label{timeline}
\vspace{-6mm}
\end{figure}

\begin{figure}
\includegraphics[scale=0.24]{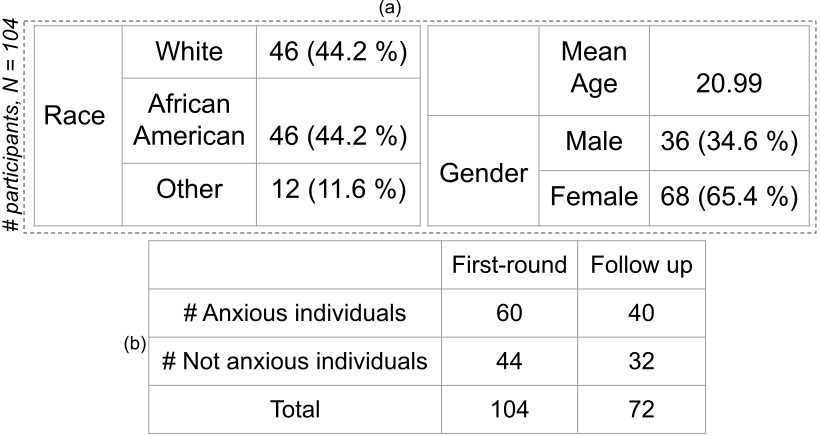}
\centering
\caption{Study population breakdown: (a) Demographics of the participants. (b) Distributions of subjects with/without anxiety conditions during the first and the follow-up rounds, computed based on the survey response via the GAD-7 questionnaire.}
\label{demographics}
\end{figure}

Five months later, we invited all $104$ participants from \textit{first-round} for follow-up and were able to follow up with $72$ individuals. We collected their Google and YouTube activity histories again, along with the survey responses for the second time. For the rest of the paper, we will refer to data collected in the second round as the \textit{follow-up} data. Therefore, there are in total $72$ people participated in both rounds and $104 - 72 = 32$ people participated only in the \textit{first-round}. The overall recruitment timeline and participant statistics are shown in Figure~\ref{timeline}. All participants were compensated with $\$10$ Amazon gift cards at the beginning during each round of participation. About $34\%$ of our participants are male and $65\%$ female. Figure~\ref{demographics}(a) presents a comprehensive breakdown of the demographics of the study population. 

\subsection{Ground Truth via Survey}\label{gt_via_survey_subsection}
\begin{figure}
\includegraphics[scale=0.5]{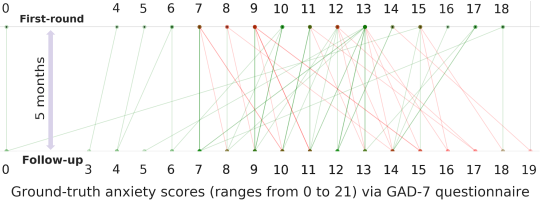}
\centering
\caption{GAD-7 scores during the \textit{first-round} and \textit{follow-up}. Red lines represent an increase, and green lines represent unchanged or decrease in anxiety scores. Multiple lines originating from one score means that there are more than one person having that anxiety score.}
\label{anxiety_dist}
\vspace{-4mm}
\end{figure}

The ground truth about one's anxiety disorder was measured using the Generalized Anxiety Disorder (GAD-7) ~\cite{spitzer2006brief}, a clinically validated questionnaire (7 questions\footnote{\url{https://www.mdcalc.com/gad-7-general-anxiety-disorder-7}}) which has been reported to be quite accurate in accessing the severity of anxiety ~\cite{swinson2006gad}. The questions in GAD-7 were prefixed with a text for the temporal context, for example, ~\textit{Over the last six months, how often have you been bothered by the following problems?} The responses were compiled to compute an anxiety score. The 21 points scale GAD-7 is a commonly used in clinical diagnosis where score of 5, 10, and 15 are treated as cutoffs for mild, moderate, and severe anxiety levels, respectively. Further follow-up and evaluation are recommended for someone with anxiety score greater than 9 ~\cite{williams2014gad}, and we used the recommended score of 9 as a cutoff to label individuals with anxiety disorder. In this work, any individual with GAD-7 score $> 9$ is labelled as \textit{Anxious}, and someone with score $\leq 9$ is labelled as \textit{Not-anxious}. Figure~\ref{demographics}(b) shows the breakdown after the anxiety cutoff. Figure~\ref{anxiety_dist} shows the distribution and changes of anxiety scores for all the participants who participated in both the \textit{first-round} and the \textit{follow-up}. We observed that the anxiety score increased for 22 individuals, decreased for 32 people, and remain unchanged for 18 participants. It is worth noticing that, $9$ participants had a change in GAD-7 score which is clinically significant (the absolute value of the change $\geq 5$) during the $5$ months of study.

\subsection{YouTube \&  Google Search History}
For this study, we collect individual-level online engagement logs from YouTube and Google Search engine using the Google Takeout interface. Google ties all online activities using the Google account associated to the user. The Takeout platform aggregates user engagement logs from all different sources and makes it available for easy accessibility. This means that as long as someone is logged into his/her/their Google account, all engagements are recorded and unified under the single Google account regardless of which device was used. For every person, the online activity history spanned (on average) over 5.7 years. In total, 1,966,400 Google searches and 1,055,847 YouTube interactions were made by all the participants.

Every engagement on YouTube and Google Search engine is timestamped along with the information whether it is the result of watching or searching. For YouTube activity logs, we use the YouTube API to extract meta-data about the videos that has been watched, which includes the title, category, video length, rating, number of likes, number of dislikes, \etc. Any video living in the YouTube ecosystem has an associated category label to it, and this enables us to get more context about the video. For Google Search activities, we label every search query text using the content classification feature of the Google Cloud NLP API \footnote{\url{https://cloud.google.com/natural-language/docs/classifying-text}}. Given a query, the API returns one or more possible category labels for the text along with a confidence score. When applicable, we select the category label with the highest confidence. The API returns a hierarchical label for every query, and we consider the root level in the hierarchy as the category label for the query. For instance, for a query $q$, if the label from the API is ``/News/Sports'', we consider ``News'' as the category for $q$. The comprehensive lists of all the categories for both search queries and YouTube videos are listed in ~\cite{googleCatList} and \cite{youtubeCatList}.


\section{Feature Extraction from Online Data}
\label{feature_engineering}
\begin{figure}
\includegraphics[scale=0.5]{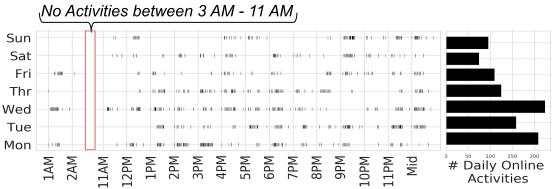}
\vspace{-6mm}
\caption{Example online activities distribution from a participant over a week, including both Google Search and YouTube activities. Each row is a day, and each `|' bar represents a single online activity. The histogram on the right side show the total online activities for each day. Notice the burstiness of daily online activities.}
\label{example_data}
\vspace{-4mm}
\end{figure}

In this section, we explain how we extracted explainable features from online history logs for each participant. Individual-level online engagement logs from YouTube and Google Search engine provides an unique opportunity to capture what may be going through one's mind at any given time. Since online activities are timestamped, one can investigate the weekday/weekend activity frequency and variance, calculate the contextual and temporal variability of these activities, and estimate daily sleeping/resting duration, \etc. For example, Figure~\ref{example_data} demonstrates the distribution of activities on YouTube and Google Search engine over a week for a specific individual in our dataset. We observe the bursty nature of incidences of these activities which we will leverage to construct features later in the section. On aggregating daily activities we found that there are higher number of interactions on these two platforms at the beginning of the week, and, as the week progresses, the number decreases. One possible explanation for drops in activities during weekends can be that people are probably spending less time interacting on internet and more time relaxing, socializing, and connecting with people around them. Notice that each of the following feature is a scalar and is calculated for each individual participant. In total, we explored five types of features, and each has a  number of variants, as described below. 


\subsection{Activity Mean and Variance}
\label{mean_var}
We define the activity mean and variance to measure the overall distribution of an individual's online interactions on YouTube and Google Search engine. We calculate the daily and weekly mean and variance of number of activities on YouTube and Google Search for each participant separately, and take the normalized $\log$ of the mean and variance for numerical stability.

\subsection{Category (\texorpdfstring{$C_{H}$}{}) \& Time (\texorpdfstring{$T_{H}$}{}) Entropy}

Every online activity has two components associated with it, namely its category and the timestamp of its occurrence. Drawing inspiration from information theory~\cite{shenkin1991information}, we define category entropy, $C_{H}$, as a measure of how diverse an individual's online activities are in terms of the semantic context. For an individual $p$, based on his/her/their online data, we compute the category entropy in the following way:
\begin{equation}
    H_{p}(Category) = -\sum_{i=1}^{m} P_{i}\times \log(P_{i})
\end{equation}
where $m$ is the number of distinct categories in the online activities of $p$, and $P_{i}$ is the percentage of activities that belong to category $i$. A high entropy indicates that $p$ interacts more uniformly across different categories online, whereas lower entropy indicates larger inequality in the number of online activities across the categories. Considering that individuals may have different habits during weekdays and weekends, we also calculated the category entropy for weekdays and weekends separately. We include the total, weekday, and weekend category entropy as features for each individual. We denote them as $C^{weekday}_{H}$, $C^{weekend}_{H}$, and $C^{total}_{H}$. 

Similarly, we define time entropy, $T_{H}$, as a measure of how diverse an individual's online activities are in terms of when it happens. We define the discrete bins for time entropy as the $24$ hours of a day. For a person $p$, time entropy is computed as below:
\begin{equation}
    H_{p}(Time) = -\sum_{i=1}^{24} P_{i}\times \log(P_{i})
\end{equation}
where the summation is taken over the $24$ hour marks, and $P_{i}$ is the percentage of activities that happen during hour $i$. A high entropy indicates that $p$ interacts with YouTube and Google Search engine more uniformly across different times of a day, whereas lower entropy indicates larger inequalities of numbers of online activities between different hours in a day. Similar to Category Entropy, we obtain the time entropy for weekdays and weekends separately. We denote them as $T^{weekday}_{H}$, $T^{weekend}_{H}$, and $T^{total}_{H}$.

\subsection{Online Activities Temporality \texorpdfstring{$\{\gamma, \alpha, \beta\}$}{}}
We observed that there is a bursty nature of online activities when plotted on the time axis (see Figure~\ref{example_data}) which resulted in clusters of online activities regardless of Google Searches or YouTube histories. In other words, we can view the incidences of online activities as a \textit{Temporal Point Process} and investigate individual-level online behaviors from a temporal point of view, such as the \textit{Inter-event Times} (IETs). We enrich our temporal feature by assuming dependencies between past activities and the next activity. The intuition is that every occurrence of an online activity increases the probability of future online activities, and the probability of the next activity decays with time. Hence, such process, called a self-exciting point process, can be modeled by the Hawkes Process ~\cite{hawkes1971spectra}, which has been widely used for modeling online data and social media activities at a population level ~\cite{rizoiu2017hawkes}. Specifically, we define a univariate Hawkes Process with an exponential decay kernel as
\begin{equation}
    \lambda(t) = \gamma + \sum_{t_i < t}\alpha \beta\exp{(-\beta(t - t_i))}
\end{equation}
where $\lambda(t)$ represents the probability (intensity) of an activity occurs at time $t$, $\gamma$ is the background intensity of an activity happens exogenously, $\alpha$ represents the ~\textit{infectivity factor} which controls the average number of new activities triggered by any past activity, and $\beta$ is the ~\textit{decay rate} where $\frac{1}{\beta}$ represents how much time has passed by, on average, between the previous event and the next event. By fitting the above Hawkes Process to each individual online history log, we obtain a unique set of $\{\gamma, \alpha, \beta\}$ for each participant as features. We keep the notations as $\{\gamma, \alpha, \beta\}$ for this set of features. 

\subsection{Inactivity Period \texorpdfstring{$\mathcal{I}$}{}}
\label{inactive}
It has been reported that YouTube is becoming the modern day classroom for students~\cite{fleck2014youtube} and provides new ways to consume contents for virtually every age groups~\cite{cayari2011youtube}. However, spending too much time on any platform can lead to internet addition~\cite{hall2001internet}, in particular the YouTube addiction~\cite{moghavvemi2017facebook} and the compulsive usage of YouTube~\cite{klobas2018compulsive}, which are quite prevalent among college population. These previous findings have inspired us consider feature that can be treated as a proxy to capture the time away from internet of each participant, and we call it the inactivity period $\mathcal{I}$. 

We focus on periods of time when no Google Search nor YouTube activity was performed of each individual. Given the online activity log of a participant and a duration threshold of $k$ hours, we pick out all the inactive periods longer than $k$ hours and investigate when they happened most frequently. Specifically, for all inactivity periods longer than $k$ hours, we first calculate the midpoint timestamp for each of them. For example, for an $8$-hour inactivity period starting at $11$ P.M. and ending at $7$ A.M., the midpoint is $3$ A.M. We found that, for all our participants and $k \in \{8, 9, 10\}$, all the midpoint modes fall in-between $5$ to $8$ A.M., which are most likely to be the middle of sleeping periods. Notice that, for the inactivity defined here, we are focusing on ~\textit{when} it occurs ~\textit{most frequently} for each individual. Hence, it is not suitable to take the mean and variance of inactivity midpoints. We included the modes of midpoints for thresholds $k \in \{8, 9, 10\}$ for each individual as features. We denote, for threshold $k \in \{8, 9, 10\}$, the inactivity mode features as $\mathcal{I}_8$, $\mathcal{I}_9$, and $\mathcal{I}_{10}$.

Overall, we developed 16 features (including variants) form the online activities (YouTube and Google Search engine) of each individual: 4 from  Activity Mean \& Variance; 3 from each of the Category Entropy $C_{H}$, Time Entropy $T_{H}$, Online Activities Temporality $\{\gamma, \alpha, \beta\}$, and Inactivity Periods $\mathcal{I}$. 

\section{Modeling Anxiety}
Following the clinical anxiety score cutoff threshold ~\cite{spitzer2006brief}, participants with GAD-7 score $> 9$ were labelled as anxious subjects, and those with score $\leq 9$ were labelled as non-anxious subjects. Overall, there were $60$ out of $104$ subjects with anxiety conditions in the \textit{first-round} and $40$ out of $72$ participants with anxiety conditions during the \textit{follow-up}. Given one's YouTube and Google Search activity history, we explore: (i) Can we identify individuals with anxiety condition through his/her/their online data? (ii) Can we predict anxiety score based on online activities and past anxiety levels? 

\subsection{Notations and Definitions}
\label{notation_section}
The feature vectors for the ~\textit{first-round} are extracted using the most recent $12$ months of data (the grey box in Figure~\ref{timeline}) before the completion of the ~\textit{first-round} survey. We denote this by $\boldsymbol{x}_1 \in \R^{16}$. Unless mentioned specifically, $\boldsymbol{x}_1$ is the concatenation of all $16$ scalar features in Section~\ref{feature_engineering} in the same order for each individual. The corresponding GAD-7 scores, gathered via the survey (the green box in Figure~\ref{timeline}) during the ~\textit{first-round}, are denoted as $y_1$. Similarly, for the ~\textit{follow-up} round, the feature vectors are extracted solely from the $5$ months of online history data (the blue box in  Figure~\ref{timeline}) ~\textbf{in-between} the ~\textit{first-round} and the ~\textit{follow-up}, and we denote it as $\boldsymbol{x}_2 \in \R^{16}$. The corresponding GAD-7 scores, provided in the ~\textit{follow-up} survey, (the magenta box in Figure~\ref{timeline}), are denoted as $y_2$. Therefore, there are in total $104$ ($\boldsymbol{x}_1, y_1$) pairs from ~\textit{first-round} and $72$ ($\boldsymbol{x}_2, y_2$) pairs from ~\textit{follow-up} (see Figure~\ref{timeline} \& Section~\ref{recruitment_procedure}).

\subsection{Classifying Individuals with Anxiety}
\label{anxiety_classification_section}
Here, we treat the problem as a binary classification task: ~\textbf{given the online activity history, we aim to identify if the subject has anxiety condition}. Assuming online activity histories are independent for every person, there are $104 + 72 = 176$ segments ($\boldsymbol{x}_1$ and $\boldsymbol{x}_2$) of online history in total, regardless of collected from which round or whom, as observation data with respective anxiety scores as labels. Formally, we are interested in $P(y \mid \boldsymbol{x})$, where $y$ is the binary anxiety label from the GAD-7 scores cutoff of $9$. 

\begin{figure}
\includegraphics[scale=0.25]{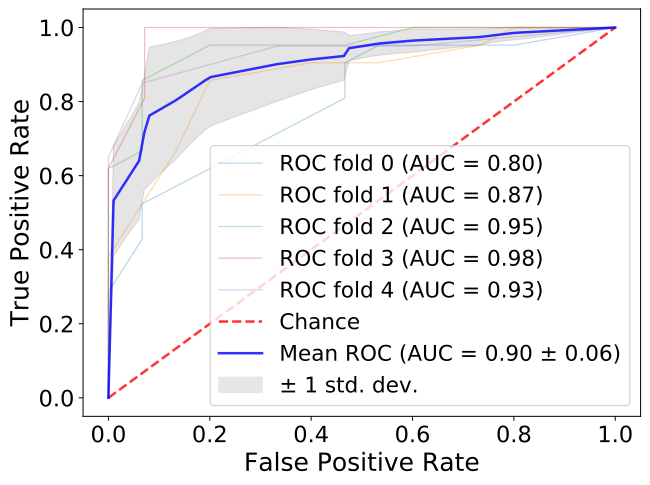}
\caption{ROC curves for Random Forests to classify individuals with anxiety. We carried out a stratified 5-fold cross-validation. The grey area represents $\pm 1$ standard deviation.}
\label{best_model1}
\vspace{-5mm}
\end{figure}

\begin{figure}
\includegraphics[scale=0.22]{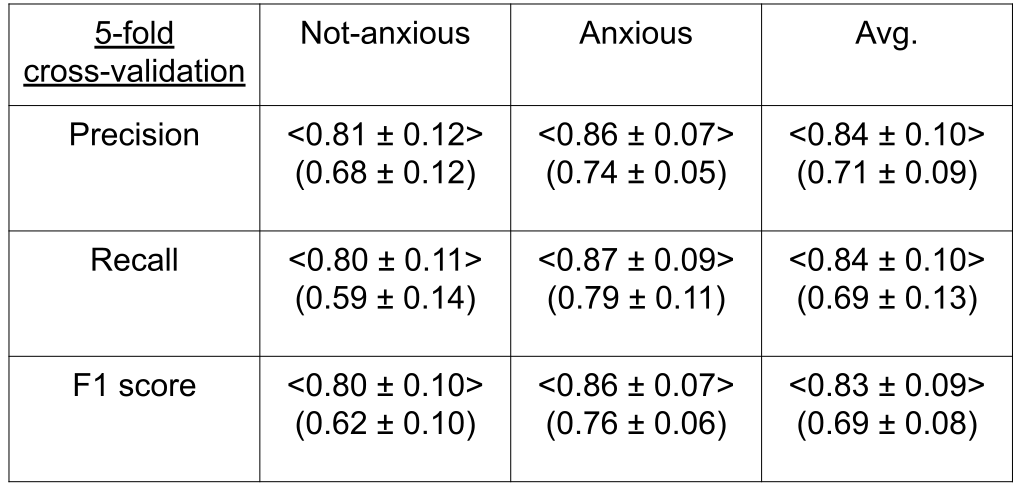}
\caption{The performance of RF and LR on the anxiety classification task. We carried out a stratified 5-fold cross-validation. The values after the $\pm$ sign represent $1$ standard deviation. Numbers inside ( ) and < > are for LR and RF, respectively.}
\label{best_model1_LR_RF}
\vspace{-6mm}
\end{figure}

We trained logistic regression (LR), linear support vector machine (SVM), and random forest (RF) classifiers on this task and performed stratified 5-fold cross-validations, respectively. Since the performances of LR and linear SVM were comparable, we report the performance of LR. However, RF significantly outperformed the other two with an average F1 score of $0.83 \pm 0.09$ and ROC AUC of $0.91 \pm 0.06$. The detailed precision, recall, and F1 scores for each class/average are reported in Figure~\ref{best_model1_LR_RF}. In Figure~\ref{best_model1}, we present the average ROC curve with standard deviations of the RF. 
\subsubsection{Possible Dependency between Two Rounds}
There are 72 out of 104 individuals who participated in both rounds of the study. Each of them has two pairs of data, $(\boldsymbol{x}_1, y_1)$ from the \textit{first-round} and $(\boldsymbol{x}_2, y_2)$ from the \textit{follow-up}. During the cross-validation for the classifiers, we manually ensured that any two pairs of data from a same participant either both belong to the training set or both belong to the testing set. We employed this to limit any personal traits or online habits, on Google Search and YouTube platform, from getting incorporated to our supervised classifiers, \IE, $\boldsymbol{x}_1 \approx \boldsymbol{x}_2$ and $y_1 \approx y_2$ for the same subject. We also experimented ordinary cross-validation without such precaution, and we observed $1 \sim 2\%$ performance increases over the metrics, indicating a small amount of potential data dependency. 

\subsection{Predicting Anxiety for Individuals}
\label{anxiety_prediction_section}
In this section, we consider the anxiety score prediction task: ~\textbf{given the online data and the past anxiety level of an individual, we aim to estimate the future GAD-7 score for that individual.} Concretely, given the two rounds of data, we aim to predict the GAD-7 score in the ~\textit{follow-up} round given the online history data \textit{and} the GAD-7 score from the ~\textit{first-round} of an individual. Formally, this task is regarded as a regression problem, and we are interested in $P(y_2 \mid \boldsymbol{x}_1, \boldsymbol{x}_2, y_1)$. 

\textit{\textbf{Features for the regression task:}} for predicting anxiety scores, $y_2$, in the above setup, we consider the weekday/weekend Time \& Category entropy $\{C^{weekday}_{H}, C^{weekend}_{H}, T^{weekday}_{H}, T^{weekend}_{H}\}$, the Temporality parameters $\{\gamma, \alpha, \beta\}$, and the Inactivity Periods with thresholds of $9$ and $10$ hours $\{\mathcal{I}_9, \mathcal{I}_{10}\}$ as input features. Thus, for the rest of the section, $\boldsymbol{x}_1, \boldsymbol{x}_2 \in R^9$ for all individuals.

We hypothesize that the change in online behaviors may preserve information about the change in anxiety level. To leverage this in the prediction task, we define the following feature vectors for the regression models:
\begin{flalign}
    & \Delta \boldsymbol{x} = \boldsymbol{x}_1 - \boldsymbol{x}_2 \in \R^{9} \\
    & \boldsymbol{x}_{gp} = [\eta \odot \boldsymbol{x}_2, (1 - \eta) \odot \Delta \boldsymbol{x}] \in \R^{2 \times 9} \\
    & \boldsymbol{x}_{reg} = [\underbrace{\eta \odot \boldsymbol{x}_2, (1 - \eta) \odot \Delta \boldsymbol{x}}_{\boldsymbol{x}_{gp}}, y_1] \in \R^{2 \times 9 + 1}
\end{flalign}
where the square bracket indicates concatenation, $\eta \in [0, 1]$ is a hyperparameter that controls the weight on $\boldsymbol{x}_2$ and $\Delta \boldsymbol{x}$, and $\odot$ denotes an element-wise multiplication. $\boldsymbol{x}_{gp}$ is a trivial modification of $\boldsymbol{x}_{reg}$ by slicing out the last entry $y_1$ and keeping only the online data features. The intuition is that $\Delta \boldsymbol{x}$ captures the shift in online behaviors between two rounds; $\boldsymbol{x}_2$ is the most recent online observation in predicting $y_2$; $y_1$ acts as a base point of $y_2$; $\eta$ weights the importance between $\Delta \boldsymbol{x}$ and $\boldsymbol{x}_2$.

\begin{figure}
\includegraphics[scale=0.3]{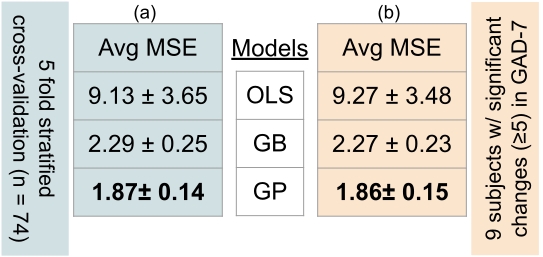}
\caption{(a) The performances of OLS, GB, and GP on the anxiety prediction task. We carried out a standard stratified 5-fold cross-validation first. (b) We then conducted another 5-fold cross-validation but kept all the 9 subjects with significant changes in GAD-7 scores in the test set. The values after the $\pm$ sign represent the standard deviations.}
\label{regression_results}
\vspace{-6mm}
\end{figure}

We chose $\eta = 0.9$ and fed the $\boldsymbol{x}_{reg}$ as inputs. For this anxiety prediction task, we first trained two models: Ordinary Least Squares regression (OLS) and Gradient Boosting regression (GB). The GB outperformed OLS significantly and achieved an average mean square error (MSE) of $2.29 \pm 0.25$ in predicting future GAD-7 scores $y_2$ (see Figure~\ref{regression_results}).


Instead of merely looking for the best prediction given by maximum likelihood estimations, it is crucial to assess the uncertainty over the model and take a Bayesian perspective, especially given we are working with healthcare applications with limited sample size. Moreover, it would grant much flexibility if the regression is not limited to parametric linear form but in a functional space with non-linearity, investigating the distribution of \textit{functions}. Therefore, we performed the regression task with a non-parametric Bayesian method, the Gaussian Process (GP)~\cite{williams2006gaussian}. We define our regression function as $f(\boldsymbol{x}_{reg})$, and it follows the GP below:
\begin{flalign}
    & f(\boldsymbol{x}_{reg}) \sim \mathcal{GP}\left(m(\boldsymbol{x}_{reg}), k(\boldsymbol{x}_{reg}, \boldsymbol{x}^{\prime}_{reg})\right)\\
    & m(\boldsymbol{x}_{reg}) = y_1\\
    & k(\boldsymbol{x}_{reg}, \boldsymbol{x}^{\prime}_{reg}) = exp\left(-\frac{\|\boldsymbol{x}_{gp} - \boldsymbol{x}^{\prime}_{gp}\|^2}{2\ell}\right)\\
    & y_2 = f(\boldsymbol{x}_{reg}) + \epsilon ~\text{where } ~\epsilon \sim \mathcal{N}(0, \sigma)
\end{flalign}
where $m(\boldsymbol{x}_{reg})$, the mean of the GP, is a deterministic function that returns the corresponding previous anxiety score $y_1$ for each subject. The covariance matrix is obtained by an exponential quadratic kernel $k$ over all pairs of individual online data, $(\boldsymbol{x}_{gp}, \boldsymbol{x}^{\prime}_{gp})$. It entails that, given any pair of individuals, the closer the distance between their online activity features in the vector space, the greater the correlation between their anxiety scores $y_2$ (close to $1$), and \textit{vice versa} (close to $0$). $\ell$ is a hyperparameter that controls the length scale between data points: the greater the $\ell$, the smoother the function. We further assume that the true $y_2$ equals to the function prediction plus an independent unknown Gaussian noise $\epsilon$, and $\sigma$ is the hyperparameter for the noise distribution. The above GP gave us a prior belief over the possible regression functions. The intuition is that, in the output space of our function $f(\boldsymbol{x}_{reg})$, the future GAD-7 anxiety scores, $y_2$, are normally distributed with a mean of the previous anxiety scores, $y_1$. The correlations between different $y_2$ values are determined by the similarities between online activities $\boldsymbol{x}_{gp}$ from the input space. 

In order to assess the performance of our GP over the test set, we first obtained the predictive posterior:
\begin{equation}
\label{pred_posterior}
    P\left(f\left(\boldsymbol{x}^{test}_{reg}\right) \mid f\left(\boldsymbol{x}^{train}_{reg}\right), \boldsymbol{x}^{train}_{reg}, \boldsymbol{x}^{test}_{reg}\right)
\end{equation}
over all the regression functions conditioned on (after observing) the training set. This conditioning operation is in a sense that, after generating functions from the GP prior, we filter out those that violate the training examples. After that, we sampled $100$ functions (traces) from the posterior from Equation~\ref{pred_posterior} and used them to make predictions on the test set. We report the average MSE of the $100$ functions, and such process is repeated for each fold of the cross-validation. We finally report the average performance over the 5 folds in Figure~\ref{regression_results}. Our GP achieved an average MSE of $1.87 \pm 0.14$ in predicting future anxiety scores $y_2$.

\textbf{Analysis:} to summarize, we evaluated the performance of our OLS, GB, and GP models for the anxiety prediction task with 5-fold stratified cross-validation. We selected 9 out of the 16 features that we extracted from online activities for this task.

Our baseline model is the OLS regression, and it assumes the anxiety score $y_2$ (dependent variable) is a linear function of the features (the independent variables). However, in reality, the relationship between anxiety and online behaviors is not necessarily linear. In order to grant more flexibility, we trained a GB regressor which consists of an ensemble of weak learners and is built in a sequential manner. The intuition is that the next learner learns from the mistakes made by the previous one, and each subsequent learner minimizes the residual prediction loss by gradient descent. The GB regressor, with an average MSE of $2.29 \pm 0.25$, significantly outperforms the OLS, whose average MSE is $9.13 \pm 3.65$, in the anxiety prediction task (see column (a) of Figure~\ref{regression_results}).

Finally, with an average MSE of $1.87 \pm 0.14$, GP outperforms both the OLS and GB (column (a) of Figure~\ref{regression_results}). Unlike OLS and GB, GP does not have any parametric form but is in a functional space with non-linearity. We assume a normal distribution for the future anxiety score $y_2$ with the \textit{first-round} anxiety score as the mean, and correlations between different $y_2$ value predictions depends on the similarities between online activities $\boldsymbol{x}_{gp}$. 

There were 9 individuals whose ground truth GAD-7 anxiety scores changed by more than $5$ between the \textit{first-round} and the \textit{follow-up}. A change in $5$ of GAD-7 scores (ranging from 0 to 21) represents a change in anxiety level by around $23$\%, which can be clinically alarming. Thus, we conducted another 5-fold cross-validation but kept all these $9$ subjects in the test set of each fold. We observed a good flexibility of $\boldsymbol{x}_{reg}$ in capturing such significant changes in GAD-7 since the performances are comparable to the average scores for all models, see Figure~\ref{regression_results}, column (b).

\section{Discussion}
It has been reported that, every year, approximately $60\%$ of all people with mental health conditions receive no treatment ~\cite{mohr2017personal}. The inability to identify patients in need of care and deliver treatments on-time are major failure points in the current healthcare system. This is mainly because our current healthcare system entirely depends on people to self-report and actively present themselves to clinics for treatments. 
Relying entirely on patients for detecting and delivering care in a timely manner is quite challenging because patients may be experiencing lack of motivations, feeling helpless and hopeless, fearing social stigmatization due to their condition, and concealing information, all of which may impair their judgements to seek help. 

In this paper, we ran a novel longitudinal study that collected ubiquitous online activities logs along with gold-standard clinically validated anxiety scores. Individual-level online activities history has been gathered from the YouTube and Google Search engine via the Google Takeout platform. We have developed explainable features that capture various semantic and temporal facets of online engagement logs, such as activity and inactivity patterns, content and time diversities. We have shown that these features are strong signals for not only detecting individuals with anxiety disorders but also estimating the severity of anxiety given any segment of online activity history. Given one's online activities, our best performing Random Forest classifier can identify an individual with anxiety condition with an average F1 score of $0.83$, average precision and recall of $0.84$, and average AUC of $0.90$. Furthermore, we have demonstrated that anxiety scores can be predicted with high accuracy, with average MSE of $1.87$, using Gaussian Process regressor.
To the best of our knowledge, we are the first to study and demonstrate that it is feasible to identify whether one is experiencing anxiety and estimate his/her/their exact anxiety score using individual-level YouTube and Google Search engine history logs. Our findings suggest the viability of constructing remote mental health surveillance frameworks based on passively sensed online data, which may be cheap, efficient, and bypasses the patient reluctance and information concealing dilemmas of traditional systems. 

\textbf{Integration into Existing Healthcare Systems:} 
the anxiety assessment framework presented in this paper can be initially set up in medical endpoints such as behavioral clinics. Therapists involved with patients suffering from various mental health conditions can use the output of the model as additional information about their patients. The predicted anxiety assessments can be leveraged to connect patients with the right counselor/expert. For example, someone may come to the clinic for a drug addiction problem and, following his/her/their informed consents, the counselor runs our model which outputs that the patient may have been experiencing severe anxiety during the last 4 week. In this case, our anxiety classification setup from Section~\ref{anxiety_classification_section} may be applicable. The patient may be flagged for review by designated members of the caregiver team who are specifically trained to handle patients with anxiety as well as addiction problems.  

Furthermore, our anxiety prediction setup from Section~\ref{anxiety_prediction_section} can be used as a guideline to initiate specific treatment steps. Most importantly, counselors can use the model on a weekly basis to monitor anxiety levels of their patients remotely (based on their online engagements) \textit{in-between} sessions/follow-up visits. This enables caregivers to note abnormal spikes in the estimated level of anxiety comparing to the last visit. Healthcare providers can then either schedule an immediate follow-up or use this information when engaging with the patient to uncover stressors and other issues that may otherwise go unmentioned during the next appointment. For example, a therapist could bring up the online behaviors of the patient during the past weekend, which were associated with high stress and anxiety symptoms, and ask if the patient agreed with the assessment. If so, what was happening in his/her/their lives at that time. Besides, such anxiety estimation setup is not one-shot fixed: it can and should be compared with professional clinical measures, as more patients came in, to help improve the future performance.

\textbf{Privacy \& Ethical Considerations:}
Building an anxiety monitoring system using individual-level YouTube and Google Search engine activity logs presents a series of concerns around privacy and data safety. Due to the sensitive nature of the data collected in this study, it is important that appropriate human subject protection protocols are in place. Hence our study protocol has been rigorously reviewed and approved by the Institutional Review Board of our institution to address these concerns. Despite these measures, we acknowledge that ethical challenges may still arise if applications based on our methods are deployed in the real world.

When someone uses platforms such as YouTube and Google Search engine, he/she/they never intend the personal data to be used by mental health assessment systems. Hence, some individuals may choose not to share their sensitive data and refuse to participate. It is important to ensure that participants, at all times, have the choice and control over their data and can choose to exclude themselves from such studies at will. Participants need to be explicitly informed about how their online engagement logs will be de-identified and analyzed, what type of information it may reveal about the user, and the accrued benefits to the patients and the therapists/care providers from mental health clinics. To address these concerns, we employed an opt-in model for volunteering study participation. In addition, we conducted one-on-one interviews for each participant during the recruitment procedure so that the research team can (a) take the time to clearly explain the purpose and the outcome of the study and (b) explicitly inform the participants about the existence of such sensitive data and how they reserve full control over the information shared such as limiting data access or deleting data. Yet, one big limitation of employing opt-in model is that it may significantly limit the number of volunteering participants for the study. Besides, the opt-in procedure may introduce participation bias in terms of study recruitment and the awareness of subjects. To limit recruitment bias, we have adapted generic wordings, such as ~``\textit{help us learn about mental health using online data}'', in our study advertisements without specifically mentioning anxiety.

Another remaining issue is that whether and when it is ethical to intervene in the life of an individual on the basis of online data signals associated with anxiety, which may not always be accurate. We believe that the ultimate decision regarding intervention must be made by therapists, care providers, and experts who understand both anxiety and the power and limitations of an automated anxiety assessment system.


\bibliographystyle{ACM-Reference-Format}
\bibliography{bibliography}

\end{document}